\documentclass[twocolumn]{aastex631}
\received{December 8, 2022}
\revised{January 30, 2023}
\accepted{January 31, 2023}
\submitjournal{ApJL}
\shorttitle{No Need for extreme stellar masses at $z\sim7$}
\shortauthors{van Mierlo et al.}
\graphicspath{{./}{figures/}}
\usepackage[super]{nth}
\usepackage{amssymb}

\defcitealias{bruzual2003}{BC03}

\begin{document}


\title{No Need for Extreme Stellar Masses at $z\sim7$: A Test-case Study of COS-87259}

\correspondingauthor{Sophie E. van Mierlo}
\email{mierlo@astro.rug.nl}

\author[0000-0001-8289-2863]{Sophie E. van Mierlo}
\affiliation{Kapteyn Astronomical Institute, University of Groningen, PO Box 800, 9700 AV Groningen, The Netherlands}

\author[0000-0001-8183-1460]{Karina I. Caputi }
\affiliation{Kapteyn Astronomical Institute, University of Groningen, PO Box 800, 9700 AV Groningen, The Netherlands}
\affiliation{Cosmic Dawn Center (DAWN), Denmark}

\author[0000-0002-5588-9156]{Vasily Kokorev}
\affiliation{Kapteyn Astronomical Institute, University of Groningen, PO Box 800, 9700 AV Groningen, The Netherlands}


\begin{abstract}

Recent controversy regarding the existence of massive ($\log(M_*/M_\odot) \gtrsim 11$) galaxies at $z>6$ poses a challenge for galaxy formation theories. Hence, it is of critical importance to understand the effects of SED fitting methods on stellar mass estimates of Epoch of Reionization galaxies. In this work, we perform a case study on the AGN host galaxy candidate COS-87259 with spectroscopic redshift $z_{\rm spec}=6.853$, that is claimed to have an extremely high stellar mass of $\log(M_*/M_\odot) \sim 11.2$. We test a suite of different SED fitting algorithms and stellar population models on our independently measured photometry in 17 broad bands for this source. Between five different code setups, the stellar mass estimates for COS-87259 span $\log(M_*/M_\odot) = 10.24$--11.00, whilst the reduced $\chi^2$ values of the fits are all close to unity within $\Delta\chi^2_\nu=1.2$, such that the quality of the SED fits is basically indistinguishable. Only when we adopt a nonparametric star formation history model within \textsc{Prospector} do we retrieve a stellar mass exceeding $\log(M_*/M_\odot)=11$. Although the derived stellar masses change when using previously reported photometry for this source, the nonparametric SED-fitting method always yields the highest values. As these models are becoming increasingly popular for James Webb Space Telescope high-redshift science, we stress the absolute importance of testing various SED fitting routines particularly on apparently very massive galaxies at such high redshifts. 
\end{abstract}


\section{Introduction} \label{sec:intro}
Over the past two decades, many examples of galaxies with stellar masses $\log(M_*/M_\odot) \gtrsim 11$ out to redshift $z\sim6$ have been found (e.g., \citealt{caputi2011,stefanon2015,deshmukh2018,marsan2022}). However, toward the Epoch of Reionization (EoR; $z\gtrsim6$), such massive galaxies become increasingly rarer (e.g., \citealt{stefanon2021}). For instance, \citet{caputi2015spitzer} found virtually no galaxy with stellar mass $\log(M_*/M_\odot) > 11.0$ at such high redshifts over 0.8\,deg$^2$ within the COSMOS field \citep{scoville2007}. This result has been recently challenged by the apparent discovery of unusually massive galaxies at $z>6$ \citep{endsley2022_alma,labbe2022}. The existence of these sources would be in tension with galaxy formation theories assuming $\Lambda$CDM cosmology (e.g., \citealt{behroozi2018,boylan2022,menci2022}). 

Galaxy stellar masses are usually estimated through spectral energy distribution (SED) fitting of photometric data. Many different SED fitting codes exists, which can lead to significantly different stellar mass estimates of the same object, especially for apparently faint galaxies (e.g., \citealt{dahlen2013,weaver2022}). Therefore, a critical study of the effects of SED fitting approaches on the derived stellar masses at $z>6$ is of utmost importance. 

For example, recent works have demonstrated that SED models assuming a nonparametric star formation history (SFH) yield higher stellar masses compared to traditional parametric descriptions \citep{tachella2022,topping2022,whitler2022}, although \citet{stefanon2022} found identical stellar masses between a constant and nonparametric SFH fit of a stacked sample of $z\sim10$ galaxies. In addition, the choice of initial mass function (IMF) also affects the derived stellar mass, and a Galactic IMF might not be the most suitable at $z>6$ \citep{steinhardt2022}.

In this letter, we present a case study of the source COS-87259, located in the third ultra-deep stripe of the UltraVISTA survey in COSMOS. This galaxy was originally identified as a $z_{\rm phot}\approx 6.6$--6.9 Lyman-break galaxy in \citet{endsley2021}. Subsequently, \citet{endsley2022_radio} identified radio and X-ray emission coming from this source, concluding that this galaxy likely harbors an active galactic nucleus (AGN). Finally, \citet{endsley2022_alma} conducted follow-up spectroscopy with ALMA, identifying strong [CII]158\,\micron\ and dust continuum emission, establishing a precise spectroscopic redshift of $z_{\rm phot}=6.853 \pm 0.002$. Endsley and collaborators obtained different stellar mass estimates for COS-87259 in their different works, with the study based on optical to far-infrared photometry including the ALMA measurement claiming that its best-estimate stellar mass is $\log(M_*/M_\odot) = 11.2 \pm 0.2$.

By using different SED fitting codes, in this work we assess whether this extremely high stellar mass value is necessarily the most accurate estimate for COS-87259. We adopt a cosmology with $H\rm{_0 = 70\ km\ s^{-1}\ Mpc^{-1}}$, $\Omega_{\rm m}=0.3,$ and $\Omega_\Lambda=0.7$. All magnitudes and fluxes are total, with magnitudes referring to the AB system \citep{okegun1983}. Stellar masses correspond to a \citet{chabrier2003} IMF. 

\section{Photometry} \label{sec:photometry}

\begin{deluxetable}{lcc}
\tablecaption{Optical and NIR Flux Density Measurements for COS-87259, as Obtained in This Work. \label{tab:flux}}
\tablewidth{0pt}
\tablehead{
\colhead{Telescope/Instrument} & \colhead{Band} & \colhead{Flux} \\
\colhead{} & \colhead{} & \colhead{($\mu$Jy)} 
}
\startdata
CFHT/MegaCam & $u$ & $<0.0039$ \\
Subaru/Suprime-Cam & $B$ & $<0.0050$ \\
Subaru/HSC & $g$ & $<0.0045$ \\
Subaru/Suprime-Cam & $V$ & $<0.014$ \\
Subaru/HSC & $r$ & $<0.0067$ \\
Subaru/Suprime-Cam & $r^+$ & $<0.013$ \\
Subaru/Suprime-Cam & $i^+$ & $<0.013$ \\
Subaru/HSC & $i$ & $<0.0079$ \\
Subaru/HSC & $z$ & $<0.0098$ \\
Subaru/Suprime-Cam & $z^{++}$ & $0.007 \pm 0.025$ \\
Subaru/HSC & $y$ & $0.097 \pm 0.053$ \\
VISTA/VIRCAM & $Y$ & $0.214 \pm 0.035$ \\
VISTA/VIRCAM & $J$ & $0.391 \pm 0.043$ \\
VISTA/VIRCAM & $H$ & $0.577 \pm 0.051$ \\
VISTA/VIRCAM & $K_{\rm s}$ & $0.837 \pm 0.081$ \\
Spitzer/IRAC & $[3.6]$ & $1.90 \pm 0.17$ \\
Spitzer/IRAC & $[4.5]$ & $1.91 \pm 0.18$ \\
\enddata
\tablecomments{For nondetections, $3\sigma$ upper limits are reported.}
\end{deluxetable}

COS-87259 is part of the UltraVISTA ultra-deep catalog presented in \citet{mierlo2022}. Based on the photometry in this catalog, our initial photometric redshift for COS-87259 is $z_{\rm phot}=6.87^{+0.08}_{-0.07}$, in excellent agreement with the spectroscopic redshift from \citet{endsley2022_alma}. To obtain more precise flux measurements for this analysis, we redid the photometry for COS-87259, using the \textsc{Python} modules \textsc{Astropy} (version 5.0.4; \citealt{astropy}) and \textsc{Photutils} (version 1.4.1; \citealt{photutils}).

We included ultra-deep optical data from Data Release (DR) 3 of the Subaru Hyper Suprime-Cam (HSC) Strategic Program \citep{aihara2022} in the $g$, $r$, $i$, $z$, and $y$ bands. In addition, we consider CFHT Megacam $u$ and Subaru Suprime-Cam broadband data, namely the $B$, $V$, $r^+$, $i^+$, and $z^{++}$ bands \citep{ilbert2009,taniguchi2015}. We also included the UltraVISTA DR4 VIRCAM $Y$,$J$,$H$, and $K_{\rm s}$ data \citep{mccracken2012} and IRAC 3.6 and 4.5 \micron\ imaging from the SMUVS program \citep{ashby2018,deshmukh2018}. In total, we consider imaging in 17 rest-frame optical to near-infrared (NIR) broad bands, which together probe the rest-frame wavelength range $450$--$6390$\,\AA\ at $z_{\rm spec} = 6.853$. 

We measured the photometry in $2"$ circular diameters at the position of COS-87259 measured from the $HK_{\rm s}$ stack, i.e., R.A. $\alpha = \mathrm{149^h 44^m 34^s.06}$ and decl. $\delta = \mathrm{+01^d 39^m 20^s.10}$. COS-87259 has only two faint low-redshift neighbors in a $5"$ radius that are both undetected in IRAC, such that we are not worried about flux contamination. Assuming a point-source morphology, aperture flux corrections were derived for each band individually, using the curves of growth of nearby bright stars, and used to correct the fluxes to total. 

To derive the flux errors in all but the IRAC bands, we measured the background fluxes in $2"$ empty apertures over a 30$"$ by 30$"$ region around the source, and calculated the flux error as the standard deviation of the flux distribution that was $3\sigma$-clipped over five iterations. As the IRAC images suffer from source confusion, for the IRAC flux errors we instead adopted a \textsc{SourceExtractor}-like approach \citep{bertin1996} meaning that we derive the flux errors from the local background measurement in a 4$"$--8$”$ diameter annulus surrounding the source.

Finally, all fluxes and error measurements were corrected for Galactic dust extinction using the \citet{schlafly2011} dust maps with the \citet{fitzpatrick1999} reddening law. In bands with nondetections, we adopt $3\sigma$ flux upper limits derived from the empty aperture fluxes. We present an overview of our flux measurements in each band in Table \ref{tab:flux}.

We compare our photometric measurements of COS-87259 to those by \citet{endsley2022_radio}, presented in their Table 1. Between the bands that overlap in both samples, our flux upper limits in nondetected bands are systematically lower, by at most a factor $\sim6$ in HSC $g$. Our VISTA $Y$, $J$, and $H$ measurements all agree within the flux uncertainty, but most importantly, our $K_{\rm s}-[3.6]$ color of 0.9 is significantly bluer than the $K_{\rm s}-[3.6] = 1.5$ from \citet{endsley2022_radio}. In Sect.\,\ref{sec:results}, we demonstrate how the photometric differences between our work and theirs affect our stellar mass comparison for COS-87259. 

\begin{deluxetable*}{lllcc}[ht!]
\tablecaption{Resulting Fit Parameters from Various SED Fitting Runs on the Optical to NIR Broadband Photometry of COS-87259. \label{tab:results}}
\tablewidth{0pt}
\tablehead{
\colhead{Code} & \colhead{Stellar Population Models} & \colhead{SFH} & \colhead{$\chi^2_{\nu}$} & \colhead{$M_*$}\\
\colhead{} & \colhead{} & \colhead{} & \colhead{} & \colhead{($\log[M/M_{\odot}]$)} 
}
\startdata
LePhare & \citetalias{bruzual2003} & Parametric & 0.66 & $10.42^{+0.22}_{-0.05}$  \\
Prospector & FSPS & Parametric & 1.11 & $11.00^{+0.05}_{-0.07}$  \\
Prospector & FSPS & Binned SFH & 2.20 & $11.16^{+0.06}_{-0.05}$  \\
LePhare & STARBURST99 & Parametric & 1.19 & $10.24^{+0.18}_{-0.05}$ \\
EAZY & FSPS/\citet{carnall22} & Parametric & 1.55 & $10.53^{+0.09}_{-0.12}$  \\
\enddata
\end{deluxetable*}

\section{SED fitting} \label{sec:sedfit}
In this section, we describe the different SED fitting approaches we took to derive the physical properties of COS-87259. We ran each code on the 17 band photometry, with the redshift fixed to the spectroscopic redshift $z_{\rm spec} = 6.853$ from \citet{endsley2022_alma}. 

\subsection{LePhare with BC03} \label{sec:lephare}
As a first code, we used the traditional, well-tested algorithm \textsc{LePhare} \citep{arnouts1999,ilbert2006}. The galaxy models were sampled from the GALAXEV library (\citealt{bruzual2003}; BC03 hereafter). 

We adopted different SFHs: a single stellar population and two parametric SFHs, i.e., an exponentially declining SFH ($\rm{SFR} \propto e^{-t/\tau}$) and a delayed exponentially declining ($\rm{SFR} \propto te^{-t/\tau}$), using star formation timescales $\tau = 0.01, 0.1, 0.3, 1.0, 3.0, 5.0, 10.0$, and 15 Gyr. We considered solar ($Z=Z_\odot$) and subsolar ($Z=0.2Z_\odot$) metallicities. 

We adopted the \citet{calzetti2000} reddening law and left the color excess free between $E(B-V)=0$--1. Emission lines were incorporated following the scaling relations from \citet{kennicutt1998} (see \citealt{ilbert2009} for a detailed description). Absorption of emission at wavelengths shorter than rest-frame $912\,\mathrm{\AA}$ by the intergalactic medium (IGM) was implemented following \citet{madau1995}. \textsc{LePhare} rejects any modeled SED that produces fluxes higher the 3$\sigma$ upper limits in the nondetected bands. 

\subsection{LePhare with STARBURST99} \label{sec:starburst99}
Young galaxies with strong nebular line and continuum emission can have significantly boosted broadband flux measurements, such that their stellar masses may be overestimated by up to a factor of 10 \citep{bisigello2019}. Therefore, we performed a separate SED fitting run with \textsc{LePhare} using stellar population models from the STARBURST99 library \citep{leitherer1999}, which include both stellar emission and nebular line and continuum emission. We considered five templates with subsolar metallicity of $Z=0.05Z_\odot$, ages spanning $10^6$--$10^8$ years, and constant SFRs between $0.01$ and $10$ $M_\odot \mathrm{yr^{-1}}$. These templates were compiled into SED models and fitted to the photometry following the \textsc{LePhare} approach described in Sect.\,\ref{sec:lephare}. 

\subsection{Prospector} \label{sec:prospector}
The second code we considered is the Bayesian inference code \textsc{Prospector} \citep{johnson2021}. This relatively new code has been extensively tested on low-redshift galaxies (e.g., \citealt{leja2017,leja2019}), but more recently it has been used in numerous works to model the properties of very high-redshift galaxies, including James Webb Space Telescope (JWST)-observed sources (e.g. \citealt{naidu2022a,tachella2022,whitler2022}). 

\textsc{Prospector} uses the Flexible Stellar Population Synthesis code (FSPS; \citealt{conroy2009, conroy2010}). We tested both a delayed exponentially declining SFH and a flexible, nonparametric SFH.

Our parametric model involves six free parameters, using the default prior shapes with the following ranges: the formed stellar mass $M_* = 10^9$--$10^{12} M_\odot$, the metallicity $\log(Z/Z_\odot) = -2$--0.19, the e-folding time $\tau_{\rm SF} = 0.001$--15 Gyr, and the age $t_{\rm age} = 0.001$--13.8 Gyr. We modeled diffuse dust attenuation following \citet{calzetti2000} with $\tau_{\rm dust} = 0$--4. Finally, we implemented IGM absorption following \citet{madau1995} and nebular emission using the default parameters.

In the nonparametric model ("continuity prior"), the SFH history is described by $N$ temporal bins, with a constant SFR in each bin, and \textsc{Prospector} fits the ratio between these bins. We largely adopted the approach outlined in \citet{tachella2022}, modeling six time bins, where the first bin spans 0--10 Myr in lookback time and the remaining five bins are spaced equally in logarithmic time space up to $z=20$. In addition, we fitted the formed stellar mass, metallicity, diffuse dust attenuation, IGM absorption factor, and gas ionization parameter, following the \textsc{Prospector} parametric model. 

Finally, \textsc{Prospector} treats nondetections by utilizing the $1\sigma$ flux limit as the flux error.

\subsection{EAZY} \label{sec:eazy}
As a third SED fitting model, we used the \textsc{Python} version of \textsc{EAZY} \citep{brammer08}. \textsc{EAZY} utilizes a series of non-negative linear combinations of basis-set templates constructed with the FSPS models. Specifically, we used the \textsc{corr\_sfhz\_13} subset of models within \textsc{EAZY}. These models contain redshift-dependent SFHs, which, at a given redshift, exclude the SFHs that start earlier than the age of the universe. The maximum allowed attenuation is also tied to a given epoch. Additionally, we included the best-fit template to the \textit{JWST}-observed extreme emission line galaxy at $z=8.5$ (ID4590) from \citet{carnall22}, which has been rescaled to match the normalization of the FSPS models. This was done to adequately model potential emission lines with large equivalent widths.

To fit our object, we adopted the \textsc{EAZY} template error function, to account for any additional uncertainty related to unusual stellar populations, using the default value of 0.2 for the template error. For nondetected bands, \textsc{EAZY} utilizes the $1\sigma$ flux upper limit in the fit.

\section{Results} \label{sec:results}

\begin{figure*}
\gridline{\fig{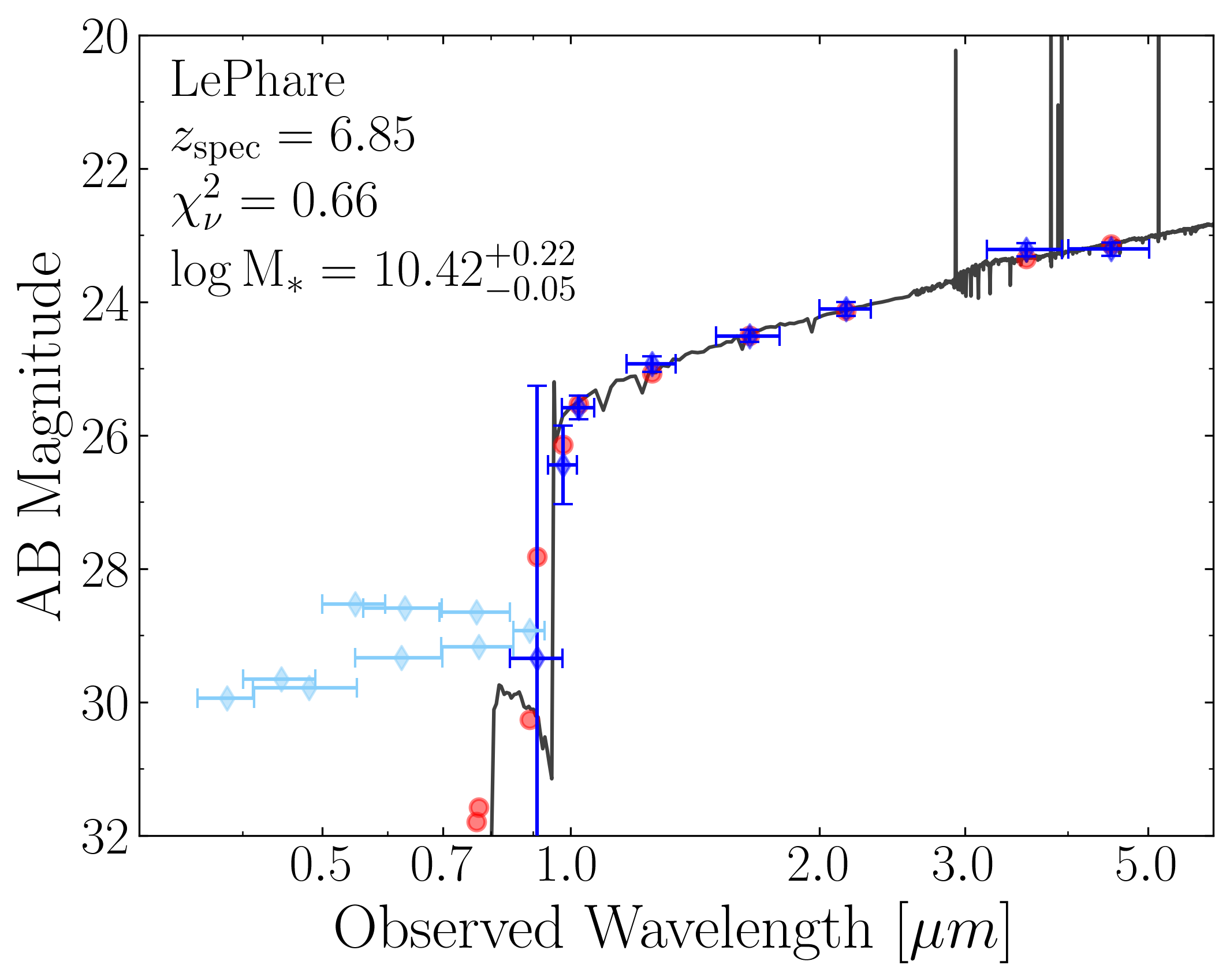}{0.5\textwidth}{(a)}
}
\gridline{\fig{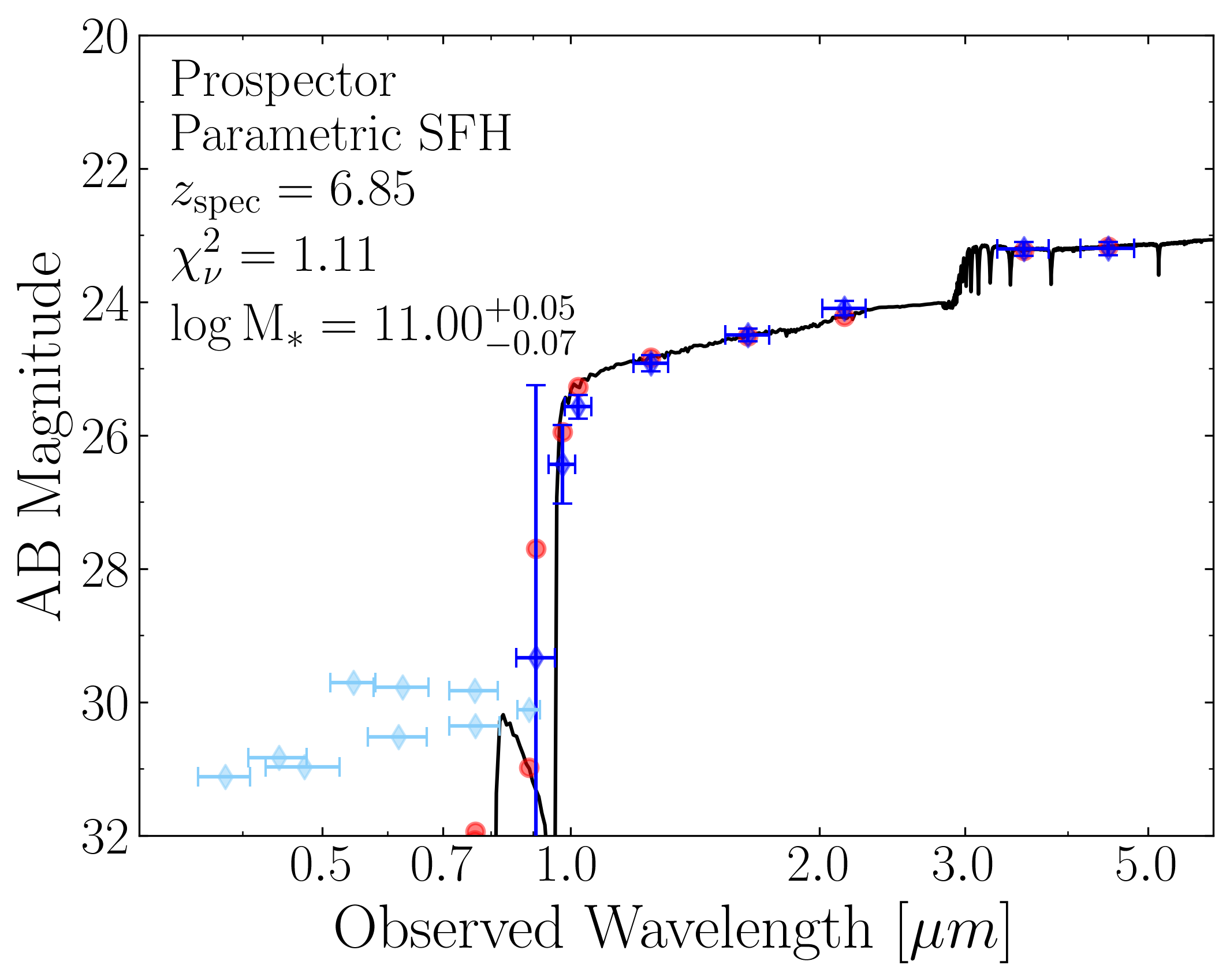}{0.5\textwidth}{(b)}
\fig{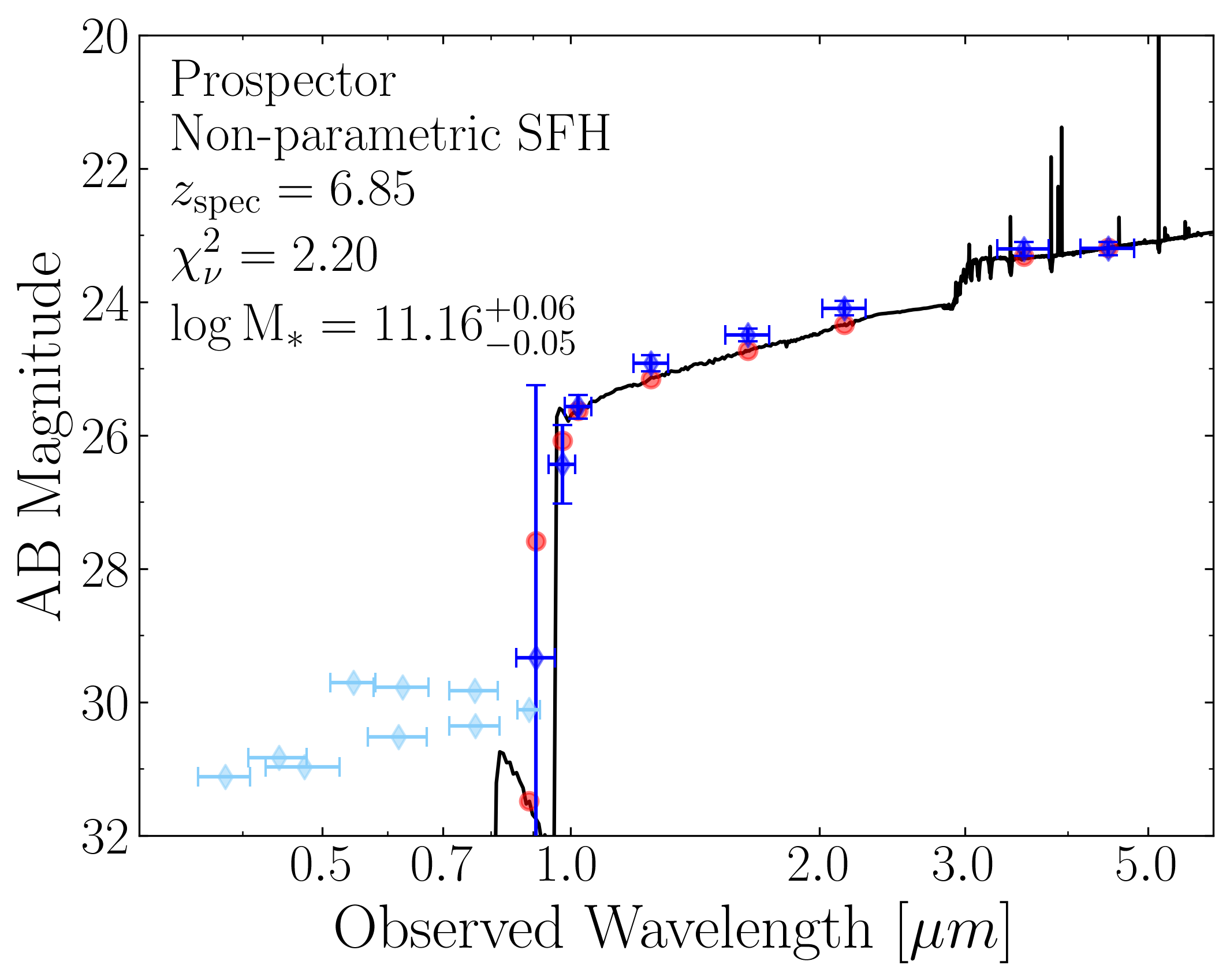}{0.5\textwidth}{(c)}
}
\gridline{\fig{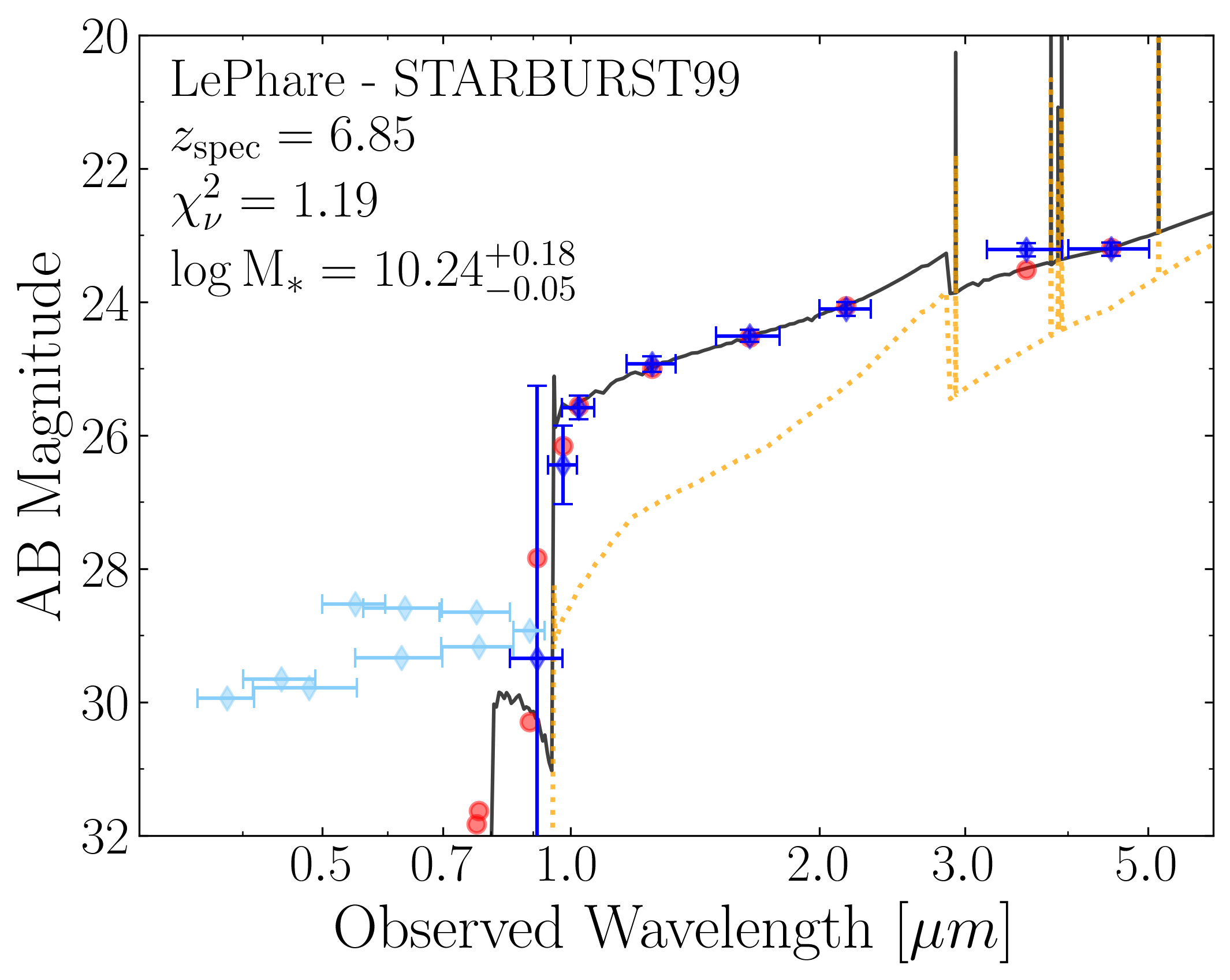}{0.5\textwidth}{(d)}
\fig{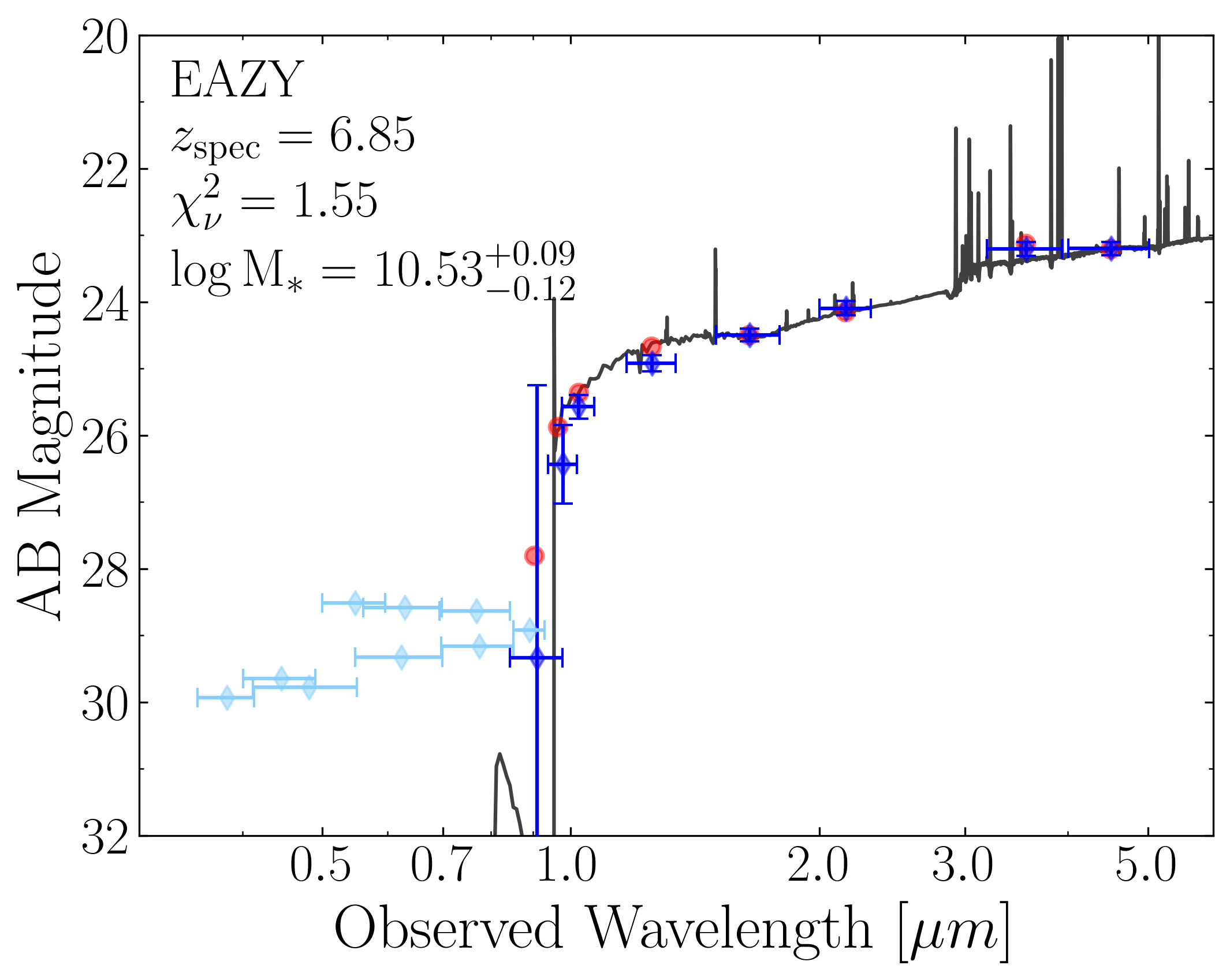}{0.5\textwidth}{(e)}
}
\caption{Best-fit SEDs corresponding to the five SED fitting setups presented in Table \ref{tab:results}. In each panel, the SED is shown in a black line. The observed fluxes and flux upper limits are shown in dark and light blue diamonds, respectively. The template fluxes in each band are shown in red points. Each panel reports the SED fitting setup, the spectroscopic redshift from \citet{endsley2022_alma}, the reduced $\chi^2$ value of the fit, and the stellar mass. For the \textsc{LePhare} run using STARBURST99 models, shown in the lower left panel, the yellow dotted line represents the contribution of nebular emission to the total SED. 
\label{fig:seds}}
\end{figure*}

Here, we compare the best-fit SEDs and associated stellar masses of COS-87259 obtained with the approaches outlined in Sect.\,\ref{sec:sedfit}. 

For each SED fitting code and choice of stellar population models, we report the reduced ${\chi^2}$ and stellar mass in Table\,\ref{tab:results}. The corresponding best-fit SEDs are shown in Fig.\,\ref{fig:seds}. Each code has its own metric to determine the best-fit SED, so we explain them as follows. With \textsc{LePhare}, the best-fit SED simply minimizes the $\chi^2$ value compared to the observed photometry, and the stellar mass uncertainties reflect the \nth{34} and \nth{66} percentiles of the maximum likelihood distribution. The resulting SEDs and stellar masses obtained with \textsc{Prospector} correspond to the median of the posterior SED, and the errors on the stellar represent the \nth{34} and \nth{66} percentiles of the stellar mass posterior distribution. Finally, \textsc{EAZY} returns the best-fit SED based on the linear combination of templates that maximizes the posterior probability distribution, with the errors on the stellar mass derived as the \nth{16} and \nth{84} percentiles. The nonhomogeneous error recipes between the codes should be taken into account account upon comparing the stellar mass errors in Table\,\ref{tab:results}, as it makes our reported \textsc{EAZY} stellar mass error inherently larger.

To compare the goodness of fit for each model, we use the reduced $\chi^2$ metric calculated from comparing the observed and modeled photometry (including contribution from emission lines) in all bands, excluding flux upper limits, where the number of degrees of freedom is simply the number of secure detections minus one, i.e., seven (see Table \ref{tab:flux}). For the stellar mass estimates presented in Table \ref{tab:results}, we imposed a minimum error of 0.05 ($10\sigma$), which was derived from the signal-to-noise ratio in the observed band that samples the SED most closely to the rest-frame $K$-band, i.e., the IRAC 4.5\,\micron\ band. At $z\sim7$, the IRAC 4.5\,\micron\ band samples the rest-frame $\sim 5500$\,\AA\ continuum and should be free from strong emission lines, such that it provides the best assessment of the stellar mass uncertainty given the our photometric data set for this source.

Our most important result is that, between the five setups of codes and stellar population models, the resulting stellar masses differ by up to 0.9 dex, whereas it is virtually impossible to determine which of these fits is most representative of the truth: the $\chi^2_{\nu}$ values are all close to one and differ by 1.54 at most. 

Between the runs performed with \textsc{LePhare} using the \citetalias{bruzual2003} models and the \textsc{Prospector} parametric setup, their best-fit SEDs shown in Fig.\,\ref{fig:seds}(a) and \ref{fig:seds}(b), respectively, the stellar masses differ by 0.6 dex and do not agree within the errors, even though we chose the run parameters to be as similar as possible. We believe this is partially because of the different templates sets, but also because of the actual SED fitting prescriptions (even if the stellar mass is derived directly from the template normalization), given that stellar mass difference between the \textsc{LePhare}+\citetalias{bruzual2003} and \textsc{LePhare}+STARBURST99 runs is only 0.18 dex.

According to the \textsc{LePhare}+\citetalias{bruzual2003} result, this galaxy would be of solar metallicity, young at 0.01 Gyr old, actively undergoing star formation with $\tau=15$ Gyr, and quite dusty with $A_V=2.03$. Instead, \textsc{Prospector} finds that this source would be metal-poor with $Z=0.014Z_\odot$, relatively old with an age of 0.18 Gyr and less dusty with $A_V=0.80$. Most importantly, its SFH e-folding time of only 0.003 Gyr and lack of nebular emission lines suggest that this galaxy would have undergone a short burst of SF upon creation and evolved relatively passively afterward. Based on the $\chi^2_{\nu}$ values of these different fits, it is impossible to say with confidence which result is more likely between these two conflicting descriptions of COS-87259's nature.

Using \textsc{Prospector}, we explicitly assess the dependency of stellar mass on the assumed SFH. We find a moderate difference of 0.12 dex in stellar mass between the two models, such that the stellar mass of $\log(M_*/M_\odot) = 11.16$ from the nonparametric SFH exceeds the parametric estimate of $\log(M_*/M_\odot) = 11.00$, even within the uncertainty. This effect has been observed in other works at similar redshift \citep{topping2022,whitler2022}. The nonparametric fit has the highest associated $\chi^2_{\nu}=2.20$, explained by the underestimation of rest-frame UV fluxes due to moderate dust attenuation of $A_V=1.44$. When we inspect our nonparametric SFH fit, we find that more than 80\,\% of the stellar mass was formed between lookback times of 0.18 Gyr and 0.042 Gyr, with a constant SFR of $1166\,M_\odot \mathrm{yr}^{-1}$. After this initial burst, the star formation rate falls off and continues at $42\,M_\odot \mathrm{yr}^{-1}$. This could explain the slightly higher stellar mass as compared to the parametric SFH, for which star formation ceases completely in the last $\sim 10^7$\,yr. 

We find that the stellar mass estimate from \textsc{LePhare} using the STARBURST99 templates is the lowest of our considered setups, at $\log(M_*/M_\odot) = 10.24$ with $\chi^2_{\nu}=1.19$. The best-fit SED is shown in Fig.\,\ref{fig:seds}(d), and corresponds to a 0.01 Gyr old galaxy with a constant SFR of $10\,M_\odot \mathrm{yr}^{-1}$ and subsolar metallicity $Z=0.05Z_\odot$. Upon decomposition of the SED, we find that 52\,\% of the integrated light is in fact nebular emission, resulting in a corrected stellar mass of only $\log(M_*/M_\odot) = 9.91$. 

Finally, we show the best-fit SED obtained with \textsc{EAZY} in Fig.\,\ref{fig:seds}(e). The stellar mass from this fit is $\log(M_*/M_\odot) = 10.53$, with an associated $\chi^2_{\nu}=1.55$, which makes the \textsc{EAZY} SED the worst fit out of the five code setups considered here. \textsc{EAZY} identifies COS-87259 as a 0.05 Gyr old galaxy, with moderate dust content, such that $A_{V}=0.87$, and a strong presence of emission lines. 

\begin{figure*}
\plotone{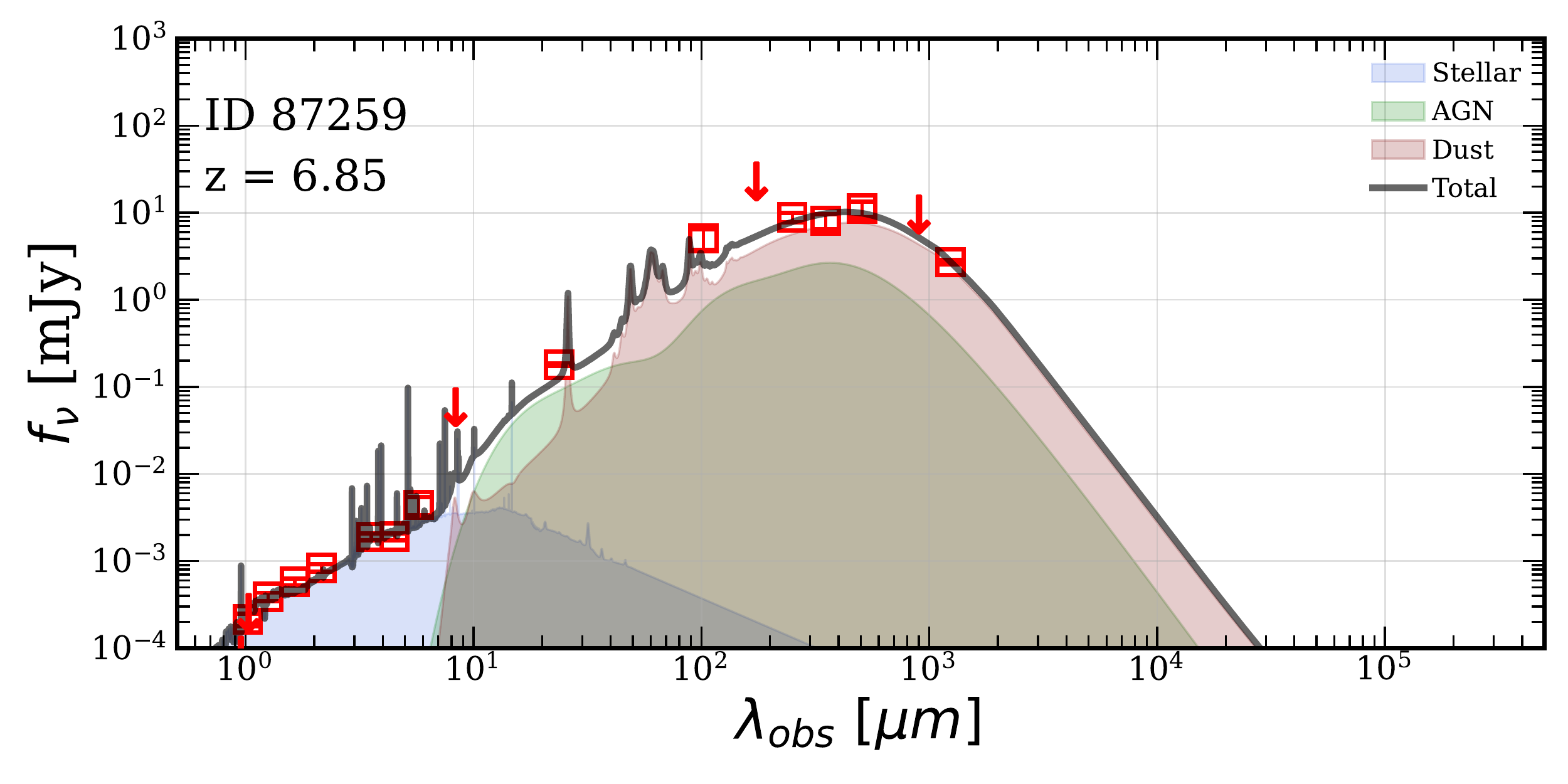}
\caption{Best-fit SED of COS-87259 obtained with \textsc{Stardust} on the observed UV to millimeter photometry from \citet{endsley2022_radio} and \citet{endsley2022_alma}. The flux measurements are shown in red, where the arrows correspond to flux upper limits. The individual components of the stellar, AGN, and dust emission are shown in blue, green, and red curves respectively. \label{fig:stardust}}
\end{figure*}

We have demonstrated how different SED fitting approaches applied to independent photometric data produce strongly varying stellar mass estimates for COS-87259. Moreover, our estimates are all lower than the $\log(M_*/M_\odot)=11.2 \pm 0.2$ value from \citet{endsley2022_alma}, although, given the uncertainty on their stellar mass estimate, it is not significantly different from our \textsc{Prospector} results. 

The stellar mass discrepancy might to some extent be explained by different photometric values: we measure a significantly bluer $K_{\rm s}-[3.6] = 0.9$ color than $K_{\rm s}-[3.6]=1.5$ from \citet{endsley2022_radio}. At $z\sim7$, the $K_{\rm s}-[3.6]$ color is sensitive to the 4000\AA\ break, but also to $\mathrm{O[III]+H}\beta$ emission which can boost the IRAC 3.6\,\micron\ flux. 

As a sanity check, we have run all codes discussed in Sect.\,\ref{sec:sedfit} on the optical and NIR photometry (HSC $g$ up to IRAC 4.5$\mu$m) from Table 1 in \citet{endsley2022_radio}, again fixing the redshift to $z_{\rm spec}=6.853$. With our \textsc{LePhare}+\citetalias{bruzual2003} setup, we retrieve a stellar mass estimate of $\log(M_*/M_\odot) = 10.87^{+0.16}_{-0.13}$. This is 0.45 dex higher than the \textsc{LePhare}+\citetalias{bruzual2003} stellar mass derived from our own photometry, and the values do not agree within the error bars. Surprisingly, the HSC IB945 flux upper limit strongly affects the stellar mass: if we change the significance of this constraint from $2\sigma$ to $3\sigma$, the best-fit SED changes to a $\log(M_*/M_\odot) = 10.58^{+0.15}_{-0.10}$ result. Furthermore, from our other code setups, we retrieve systematically higher stellar masses for COS-87259 compared to our own results, but the same spread in masses of 0.9 dex. Therefore, we conclude that our lower stellar mass estimate for COS-87259 is partially driven by our photometry, but the discrepancies between the results of our considered SED fitting routines are not. 

\section{Combined stellar and dust emission SED fitting} \label{sec:discussion}

Another key difference with respect to our analysis that could explain the $\log(M_*/M_\odot)=11.2 \pm 0.2$ result from \citet{endsley2022_alma} is the inclusion of FIR to millimeter data, especially given that COS-87259 likely harbors an AGN. In fact, \citet{endsley2022_alma} have fitted the full optical to millimeter wavelength photometry with a custom Bayesian SED fitting package, which includes AGN and galaxy dust emission components, as well as stellar emission using the FSPS code under the \textsc{Prospector} framework. So far, we have only fitted the optical to NIR regime, so here we adopted the 5.8\,\micron\ to 1.4\,mm photometry for COS-87259 from \citet{endsley2022_alma} and Table 1 in \citet{endsley2022_radio}.

This combined suite of photometry is fitted with the SED-fitting code \textsc{Stardust} \citep{kokorev21}, again fixing the redshift to $z_{\rm spec}=6.853$. \textsc{Stardust} models light from stars and AGN, as well as infrared emission arising from the dust reprocessed stellar continuum. Similarly to \textsc{EAZY}, \textsc{Stardust} fits independent linear combinations of templates, but with the key advantage of not assuming the energy balance between stellar and dust emission. For our fit, we utilized UV-NIR templates adopted from \textsc{EAZY}, empirically derived AGN templates from \citet{mullaney11}, and the dust models from \citet{draine07}. For the latter, the minimum radiation fields intensity spans $U_{\rm min}=40$--50, and the fraction of dust contained in the photodissociation regions spans $\gamma=0.01$--0.3. When combined, these correspond to a range of luminosity-weighted dust temperatures ($T_{\rm dust}$) from 35 to 45 K. 

The best-fit SED obtained from \textsc{Stardust} is shown in Fig.\,\ref{fig:stardust}. We note that \textsc{Stardust} treats any flux measurement with a confidence level $<3\sigma$ as an upper limit instead, which brings the the total number of secure detections to 14. The $\chi^2_{\nu}$ value for this fit is 1.90, and the resulting stellar mass is $\log(M_*/M_\odot)=10.81^{+0.05}_{-0.05}$, which is over 0.4 dex lower than the $\log(M_*/M_\odot)=11.2 \pm 0.2 $stellar mass from \citet{endsley2022_alma} and does not agree within the error bars. When we run \textsc{Stardust} on the exact photometry from \citet{endsley2022_radio}, we retrieve a stellar mass of $\log(M_*/M_\odot) = 11.02^{+0.05}_{-0.05}$. These results with \textsc{Stardust} reinforce our previous conclusion obtained with the SED fitting of only the source stellar emission: the stellar mass of COS-87259 is most likely $<10^{11.2}\, M_\odot$.

\section{Conclusion}\label{sec:conclusion}
In this letter, we reassessed the stellar mass of $z_{\rm spec}=6.853$ AGN host galaxy candidate COS-87259, located in the UltraVISTA ultra-deep region in the COSMOS field. This source has been extensively studied in previous works: its most recent stellar mass estimate is unexpectedly high, with $\log(M_*/M_\odot) = 11.2 \pm 0.2 $ \citep{endsley2022_alma}. Here, we took this galaxy as a case study to compare the best-fit SEDs and physical parameters obtained with different SED fitting routines. We measured independent photometry from 17 rest-frame optical to NIR broadband images for COS-87259. These data were fitted with SED fitting codes \textsc{LePhare}, \textsc{Prospector}, \textsc{EAZY}, and \textsc{Stardust}, including 5.8\,\micron\ to 1.4\,mm photometry from \citet{endsley2022_alma} for the latter fit. 

Between six setups of codes and stellar population models, we find that the resulting stellar masses span $\log(M_*/M_\odot)=10.24$--11.16. Contrarily, the reduced $\chi^2$ values of the fits are all close to unity within $\Delta \chi^2_\nu=1.2$. Therefore, all SED fits are of comparable quality, making it virtually impossible to decide which stellar mass estimate is most representative of the truth. 

We find that the combination of \textsc{Prospector} and a nonparametric description of the SFH (which has been frequently used to fit newly \textit{JWST}-discovered high-redshift galaxies) yields the highest stellar mass estimate in this work, $\log(M_*/M_\odot)=11.16$. Moreover, even when adopting a traditional parametric SFH, \textsc{Prospector} yields significantly higher stellar masses than any of the other considered codes. Finally, by considering very young galaxy templates that have strong nebular line and continuum emission, we obtain our lowest stellar mass estimate of $\log(M_*/M_\odot)=10.24$ with $\chi^2_{\nu}=1.19$. 

We emphasize that none of our six considered SED fitting routines can replicate the extremely high stellar mass result from \citet{endsley2022_alma}, although this is partially explained by our bluer $K_{\rm s}-[3.6]$ color measurement. It should be noted however that a $\log(M_*/M_\odot) \geq 11$ solution for COS-87259 does not violate $\Lambda$CDM number density upper limit \citep{boylan2022}, even when other $z\sim7$ galaxies of such stellar masses may be discovered in the COSMOS field in the future \citep{lovell2023}. 

In conclusion, in light of the recent discoveries of very massive EoR galaxies with \textit{JWST}, we emphasize the absolute importance of testing various SED fitting routines on these seemingly massive galaxies to obtain a confident stellar mass estimate. Otherwise, we may falsely conclude that \textit{JWST} is allowing us to probe an unexpectedly numerous population of massive galaxies, whereas in fact overestimation from novel SED fitting approaches is the main driver behind these results. As for the specific instance of COS-87259, this source will be observed with \textit{JWST} in the near future, hopefully bringing us yet again closer to a consensus on the nature of this undoubtedly interesting galaxy. 

\begin{acknowledgments}
We thank Ryan Endsley, Joel Leja, Chris Lovell, and Mara Salvato for useful discussions. We are also grateful to the anonymous referee for a careful reading of the manuscript and a constructive report. K.I.C. and S.v.M. acknowledge funding from the European Research Council through the award of the Consolidator grant ID 681627-BUILDUP. K.I.C. and V.K. acknowledge funding from the Dutch Research Council (NWO) through the award of the Vici grant VI.C.212.036.
\end{acknowledgments}

\end{document}